\documentclass[12pt]{iopart}

\usepackage{iopams}  
\usepackage{graphicx}
\begin{document}

\title[]{ Spin Entanglement in supramolecular structures.}

\author{F. Troiani$^1$, V. Bellini$^1$, A. Candini$^1$, G. Lorusso$^{1,2}$ and M. Affronte$^{1,2}$.}

\address{$^1$CNR-INFM-S$^{3}$ National Research Center on nanoStructures and bioSystems at Surfaces \\
$^2$Dipartimento di Fisica, Universit\`a di Modena e Reggio Emilia, via Campi 213/a, 41125 Modena, Italy;\\}

\ead{marco.affronte@unimore.it}
\begin{abstract}

\end{abstract}

Molecular spin clusters are mesoscopic systems whose structural and physical 
features can be tailored at the synthetic level.
Besides, their quantum behavior is directly accessible in laboratory and their 
magnetic properties can be rationalized in terms of microscopic spin models. 
Thus they represent an ideal playground within solid state systems to test 
concepts in quantum mechanics. One intriguing challenge is to control
entanglement between molecular spins. Here we show how this goal can be 
pursued by discussing specific examples and referring to recent achievements.

\maketitle

\section{Introduction}

Entanglement is a peculiarity of quantum systems and it represents one of the 
most fascinating aspects of quantum mechanics. It essentially consists in the 
impossibility of describing a quantum object without some knowledge on the rest 
of the system.
More formally, it expresses the impossibility of factorizing the wavefuncion 
of a composite system into the product of the wavefunctions of the components. 
For photons or cold atoms, as well as for few solid state systems, entanglement 
is largely investigated, both theoretically and experimentally \cite{Horodecki, 
fazio}.
These achievements underpin and stimulate exploitation of this property for new applications like quantum cryptography, teleportation and computation. Besides,
the controlled generation of entanglement between nanoscaled objects allows to 
explore the boundary between quantum and classical behaviour. 

Molecular spin clusters represent a very interesting test bed in this context.
In fact, they represent complex but finite systems whose structural and physical features can be tailored at the synthetic level and whose collective properties 
can be predicted by microscopic, albeit demanding, models. Recent achievements 
on supramolecular chemistry, experiments and modeling appear extremely 
encouraging in this field.\\
Here, we briefly review suitable molecules and linkers and illustrate methods 
used for the experimental determination and rationalization of supramolecular 
systems. With the help of specific examples, we discuss different issues 
including the possibility of quantifying and probing entanglement in 
supramolecular systems; besides, we provide hints to understand and control 
the inter-molecular coupling.

\begin{figure}[ptb]
\begin{center}
\includegraphics[width=15cm]{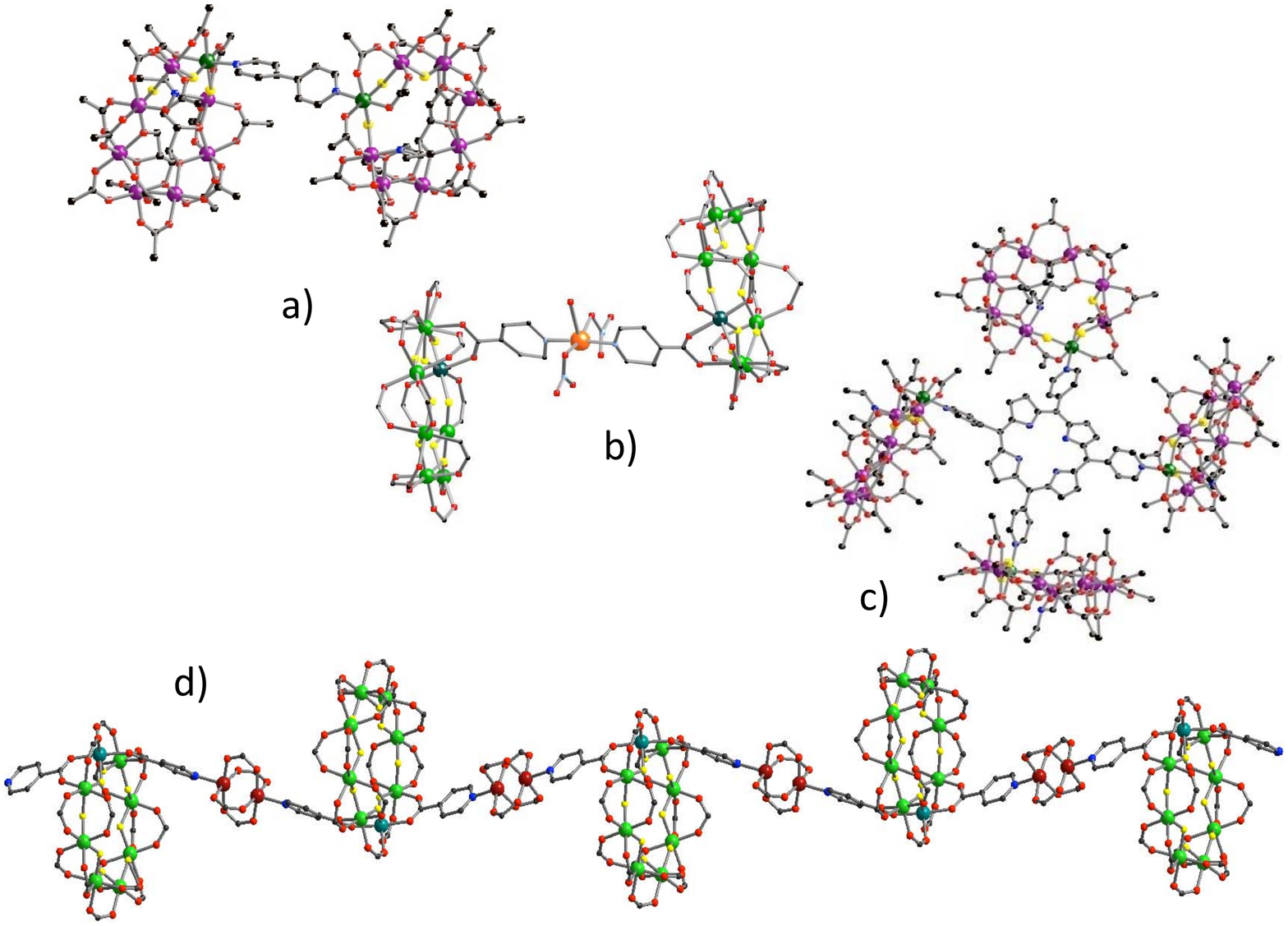}
\end{center}
\caption{Supramolecular structures based on Cr$_7$Ni rings. $a)$ two \textit{purple}
Cr$_7$Ni rings linked by bipyridine \cite{AngewChemGlued}; 
$b)$ two \textit{green} Cr$_7$Ni rings linked by a metallorganic group containing a
metal ion \cite{NanoNature}; $c)$ a tetramer formed \textit{purple} Cr$_7$Ni rings
\cite{AngewChemGlued}; $d)$ chain alternating Cr$_7$Ni rings with Cu ($s$=1/2) ions.}
\label{structures}
\end{figure}

\subsection{Molecular spin clusters}

Molecular spin clusters are molecules consisting of a magnetic core and an external non-magnetic shell. Typically, the inner part is made of transition metal (hydro-)
oxides bridged and chelated by organic ligands (typically chemical groups comprising 
light elements like carbon, oxygen, hydrogen, nitrogen, etc.). Once synthesized, 
magnetic molecules are generally stable and they can be dissolved in solutions.
From these, bulk crystals, comprising a macroscopic number of identical units aligned 
along specific crystallographic directions, can be obtained. In general, molecules 
are not interacting with each other and the behavior of a bulk crystal turns out to 
be that of a collection of non-interacting, identical molecules.
This allows to use conventional solid state experimental techniques to investigate 
molecular features, which is certainly one of the keys for success of these molecular objects. In the recent years, part of the interest in the field has turned at developing protocols to graft and study arrays of molecules on suitable substrates, aiming at addressing few or - eventually - single units. \\
Within each molecule, uncompensated electron spins are well localized on transition 
metals with quenched orbital moments (Fe, Mn, Cr, Ni, Cu...) and interact with each other by (super-)exchange coupling.
These ferro- or antiferro-magnetic coupling dominates the intramolecular interactions and determines the pattern of magnetic eigenstates. Typically, the molecular spectra are well resolved at liquid-helium temperatures, while multiple level crossings can be observed at magnetic fields of few Teslas, that are easily achieved in laboratory. Anisotropy and antisymmetric terms in the spin Hamiltonian of the single molecule may arise from reduced local symmetries.\\ 
In the last years, most of the interest has been devoted to molecules like the prototypical Mn$_{12}ac$ or Fe$_8$, with high-spin ground state and high anisotropy barrier, that exhibit a characteristic hysteresis loop of the magnetization, 
justifying the name of single molecule magnets (SMM) \cite{mmagnets}.
Intermolecular interactions can be reduced by diluting molecules in solid crystals \cite{Ga6, Fe18} or in frozen liquid solution. Intermolecular dipolar interaction is limited in the case of antiferromagnetic molecular clusters, characterized by 
low-spin ground states. Among these, molecules with S=1/2 ground state, like V$_{15}$ \cite{V15a, V15b} or the heterometallic Cr$_7$Ni rings \cite{QC1} represent prototypical examples of mesoscopic effective two-level systems.\\
A relevant aspect is the coherence of the molecular spin dynamics. Generally speaking, SMM represent an ideal playground to observed quantum phenomena at mesoscopic scale \cite{mmagnets}.
The spectral definition of the SMM ground multiplet allowed to perform electron spin resonance experiments in Fe$_8$\cite{pulsedFe8}, Ni$_4$ \cite{Ni4} and Fe$_4$ 
\cite{pulsedFe4}; these capabilities inspired schemes for performing quantum 
algorithms in Mn$_{12}ac$ or Fe$_8$ \cite{grover}, based on the massive exploitation 
of linear superpositions and quantum intereference.
A special case of coherent spin dynamics is that observed in single rare earth ions diluted in a crystalline matrix \cite{pulsedEr}, that, however, do not represent a mesoscopic system.
More recently, time resolved experiments have shown that molecular electron spins can be coherently manipulated. In the case of antiferromagnetic clusters, Rabi 
oscillations in the 10$^{-1}\,\mu$s time scale have been observed in V$_{15}$ while decoherence time $\tau_d$ as long as  3$\,\mu$s at 2K have been directly measured in molecular Cr$_7$Ni rings \cite{AA}.
Since the gate time $\tau_g$ to manipulate the effective S=1/2 in real experimental conditions is of the order of 10$\, ns$, it turns out that the figure of merit $Q=\tau_d/\tau_g$ exceeds 100 at 2K for Cr$_7$Ni.
For an isolated molecule the main source of decoherence remains the interaction with the nuclear spins both at the metal sites (specific isotopes) or in the organic environment (protons, fluoride, etc.). Molecules typically comprise few hundreds of atoms in well defined positions, so the interactions between the electron and the nuclear spins can be rationalized for each molecule \cite{deco}.

\subsection{chemical routes for linking molecules}

Entangling spins in supramolecular structures, such as nanomagnet dimers or oligomers, requires at least two separate steps:
1) the individuation of molecular building blocks with well defined features;
2) the establishment of inter-molecular magnetic coupling.  
Concerning the first step, the synthesis
and the characterization of separate molecular units should be considered as prerequisite.
Ideally, each of the molecular units should be individually addressable; 
this implies that they
should be either spatially or spectrally resolvable.\\
Different kinds of magnetic coupling between the units are compatible with the controlled generation of entangled states.
Dipolar interaction is long-range and might be desirable if one searches entanglement of a large collection of objects \cite{Ghosh}, but it is detrimental to control
entanglement between few molecular units within an oligomer, for 
it tends to couple molecules belonging to different oligomers. Therefore, local types of magnetic interaction, such as exchange, are preferable. In practice, when organic linkers are used to exchange couple magnetic molecules there
are two main risks: 1) to form polymeric networks that tend to undergo long range magnetic order; 
2) magnetic states of the single moiety can be heavily perturbed by the chemical link. 
Recently different
aromatic groups have been successfully used to selectively link molecular spin clusters. G. Timco and
R.E.P. Winpenny in Manchester are currently using piridyne and pyrazole groups \cite{AngewChemGlued}
while the group of G. Aromi is using $\beta$-diketonates ligands \cite{aromiMn4, aromiCuNi}.

Probably the first case of molecular dimer reported in the literature is the 
[Mn$_4$]$_2$ \cite{Mn4WW,Mn4SH}. The individual moiety, 
[Mn$_4$O$_3$Cl$_4$(O$_2$CEt)$_3$(py)$_3$]\cite{Mn4} comprises three Mn$^{+3}$ and one Mn$^{+4}$ coupled together to give a S=9/2 ground molecular state and a uniaxial anisotropy.
Two Mn$_4$ are linked through hydrogen bonds to forms Mn$_4$ dimer in which the magnetic states of each moiety are antiferromagnetically coupled to each other. 
The true problem of entanglement however was not considered there.\\
Another important case is that of heterometallic Cr$_7$Ni rings. Two species of Cr$_7$Ni rings have been synthesized: \textit{green} \cite{Cr7M} and \textit{purple}\cite{AngewChemGlued} Cr$_7$Ni, after their respective colour. The first attempt of 
linking two \textit{green} Cr$_7$Ni rings was through the internal amine and 
different metallorganic groups \cite{AngewChemCr7Ni}. From the chemical point of view this was successfull since two rings have been selectively linked. Yet, the magnetic coupling
resulted vanishingly small except in the case where a Ru$_2$ dimer was introduced in the linker \cite{Ru2}.
That was interesting since this Ru$_2$ dimer has redox properties and in principle its magnetic features
can be switched by an external electrical stimulus; however, the effectiveness of such a scheme still
needs to be proved. Important progress have been recently obtained exploiting the fact that the chemical
reactivity of the extra Ni is much faster than that of the rest of the Cr ions in the rings. Firstly, a chemical
group was attached to the carboxylate at Ni site in the \textit{green} Cr$_7$Ni \cite{NanoNature}; more
recently, nitrogen of heterocyclic aromatic groups was directly linked to the Ni in the \textit{purple} Cr$_7$Ni
\cite{AngewChemGlued}. Starting from these, the choice of the linker is virtually infinite \cite{Timco}.
In a first series of linked \textit{green} Cr$_7$Ni rings, transition metal ions (M) or dimers were inserted
in the linker thus forming Cr$_7$Ni-M$_{x}$-Cr$_7$Ni  with x=1,2 \cite{NanoNature}.
By using \textit{purple} Cr$_7$Ni, a family of [Cr$_7$Ni]$_2$ with short or longer linkers was obtained,
thus allowing to tune the strength of the intramolecular coupling \cite{AngewChemGlued}. This strategy
can also be used to synthesize molecular trimers, tetramers (with or without central metal ions) or chains
alternating Cr$_7$Ni and metal ions or dimers (see Fig. \ref{structures}) \cite{AngewChemGlued,Timco}.

\subsection{measuring and quantifying the magnetic coupling}

\begin{figure}[ptb]
\begin{center}
\includegraphics[width=10cm]{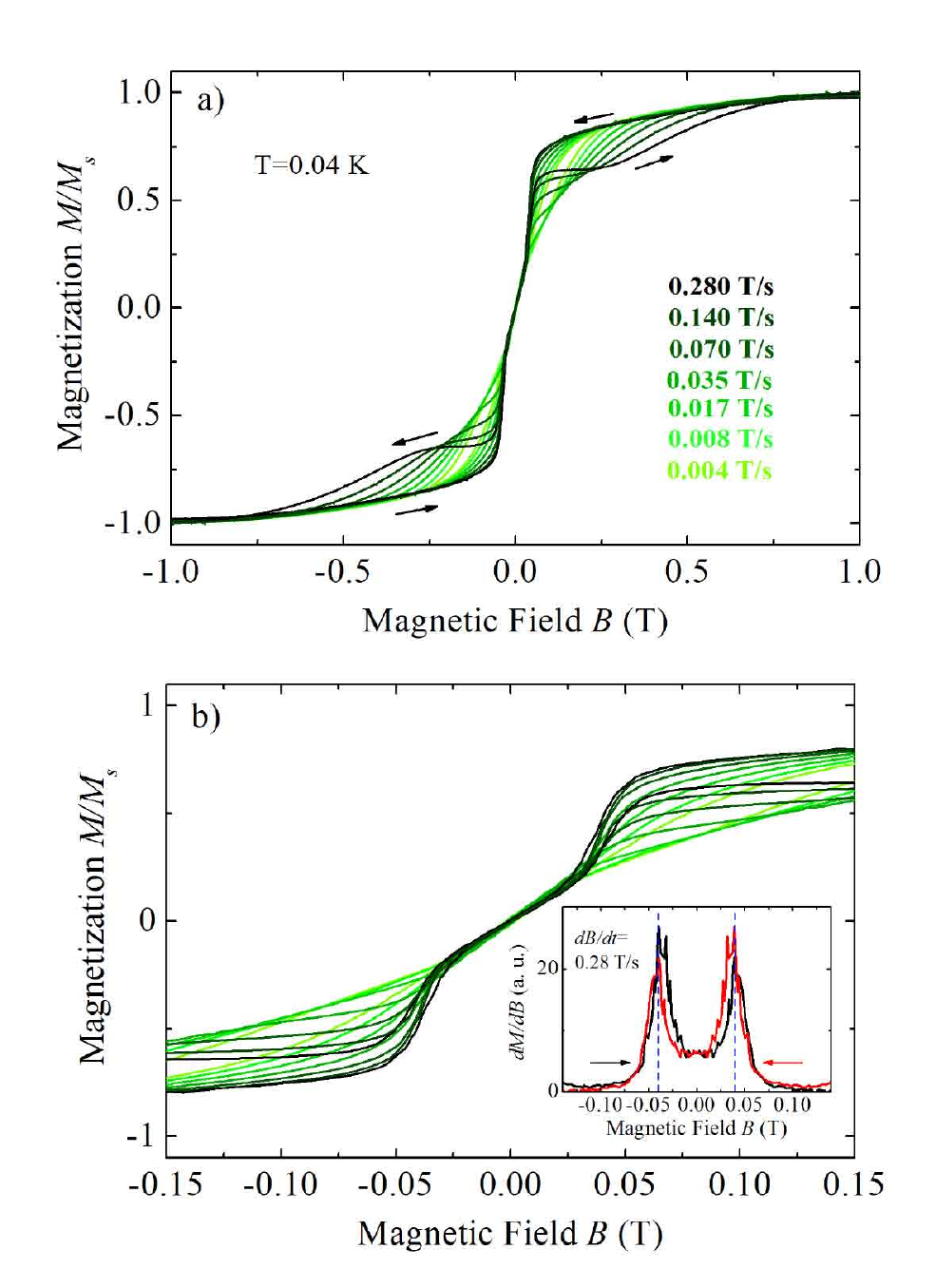}
\end{center}
\caption{ Experimental magnetization curves M(T, B)  taken for [Cr$_7$NiF$_3$(Etglu)(O$_2$CtBu)$_{15}$]$^{2-}$
(dipyet) (Etglu=N-ethyl-d-glucamine and dipyet= trans-1,2-dipyridylethene).
a) Data are taken at T=40 mK and different sweeping rates of the magnetic field  \cite{ent09}.
b) Magnification of a). (Inset) dM/dB vs B curve taken for dB/dt=0.28 T/s.}
\label{magnetization}
\end{figure}

The magnetic effectiveness of the intramolecular link can be experimentally evaluated.
According to what was previously discussed, we firstly check the integrity of each molecular sub-unit
and then we quantify the strength of the coupling. This may require the use of complementary
experimental techniques and, possibly, the systematic comparison within a series of derivatives,
from the individual molecule to complex aggregates. Magnetic susceptibility and magnetization loops
are primarily used to clarify the nature of the ground state of the system while specific heat measurement
directly evaluates the energy splitting of the lowest multiplets. Both need to be extended to very low
temperatures (typically T$\, < \,$1K) where the magnetic coupling becomes observable. Electron paramagnetic resonance (EPR)
spectra allow to evidence transitions that are permitted only when the magnetic coupling is effective and they are
sensitive to the anisotropy of the $g$-factor.\\
As an example, Figure \ref{magnetization} shows the magnetization loop $M(T,B)$ for a
[\textit{purple}-Cr$_7$Ni]$_2$ dimer with a trans-1,2-dipyridylethene ligand between two rings \cite{ent09}.
The $M(T,B)$ curves presented in the upper panels show the butterfly behaviour, typical of the
phonon-bottleneck regime, that becomes clearer as the sweeping rate dB/dt increses.
Zooming the magnetization curves $M(B)$ (lower panel), we can observe the presence of fleeble knees, that are clearly evident by taking the derivative of magnetization  $dM/dB$ as shown in the insets. 
These features are not present in the single purple-Cr$_7$Ni ring and they are clearly due to the intra-molecular coupling.

\begin{figure}[ptb]
\begin{center}
\includegraphics[width=10cm]{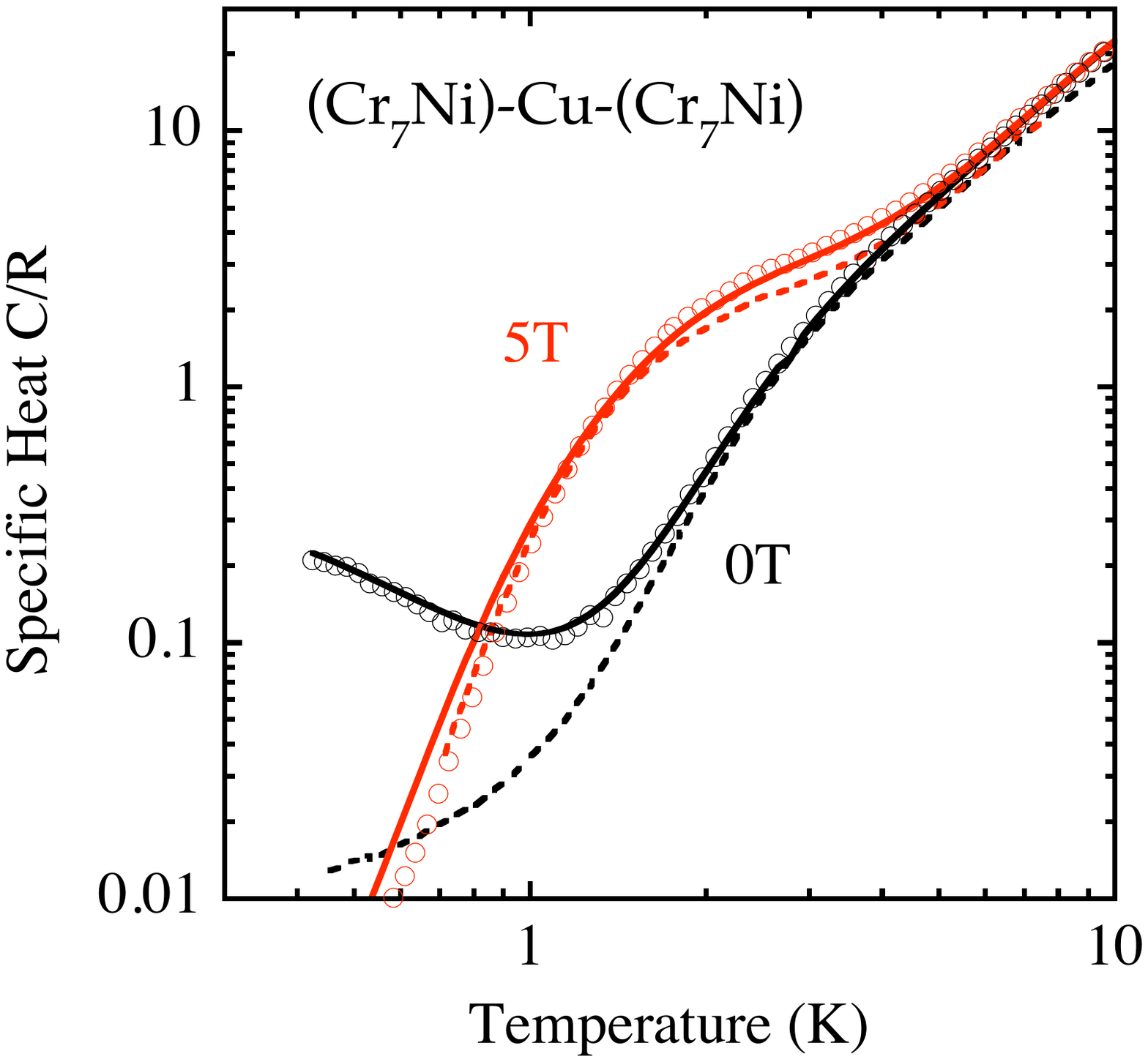}
\end{center}
\caption{Low temperature specific heat of Cr$_7$Ni-Cu-Cr$_7$Ni molecular trimer
(circles). The -experimental- specific heat of two individual rings per unit cell is plotted as dotted lines.
Continuos lines are calculated by spin hamiltonian (see text) and they perfectly reproduce the experimental data.}
\label{HC}
\end{figure}

In Fig. \ref{HC} we consider another typical case,  the Cr$_7$Ni-Cu-Cr$_7$Ni molecular trimer,
for which the specific heat C(T) provided direct evidence and quantification of the supramolecular
coupling \cite{NanoNature}. This system comprises two Cr$_7$Ni rings with an S=1/2 ground state doublet and an S=3/2 first excited multiplet, and one Cu ion with S=1/2. The bumps in the C(T) curve
are the Schottky anomalies related to the energy splitting of specific multiplets.
In 5T the main anomaly is essentially related to the splitting
between the S=1/2 and S=3/2 multiplets, typical of the individual Cr$_7$Ni.
The overlap between the specific heat of Cr$_7$Ni-Cu-Cr$_7$Ni (circles in Fig. \ref{HC}) and that
of two times the C(T) of individual Cr$_7$Ni rings (dotted lines in Fig. \ref{HC}) is a direct evidence of the integrity
of the molecular rings. In zero field, a Schottky anomaly clearly appears below 1K for
Cr$_7$Ni-Cu-Cr$_7$Ni but it is not present for individual rings for which the ground state is
a Kramer doublet. This low temperature anomaly is a consequence of the coupling between
the three effective spins S=1/2 in Cr$_7$Ni-Cu-Cr$_7$Ni.

It's worth stressing the sophisticated level of description of these mesoscopic systems provided by microscopic spin Hamiltonians. Briefly, the spin Hamiltonian of a single Cr$_7$Ni ring reads:
\begin{eqnarray}
{\cal H} & = & 
J \sum_{i=1}^8  {\bf s}_{i} \cdot {\bf s}_{i+1} +
\sum_{i=1}^8 d_i\, [s_{z,i}^2-s_i (s_i +1)/3] \nonumber \\
& + &
\sum_{i<j=1}^8 {D}_{ij} [2 s_{z,i} s_{z,j}-s_{x,i} s_{x,j}-s_{y,i} s_{y,j}]
+ \mu_B \sum_{i=1}^8 {\bf B}\cdot {\bf g}_i \cdot {\bf s}_{i} ,
\label{eq1}
\end{eqnarray}
where the $z$ axis coincides with the ring axis, site 8 corresponds to the Ni$^{2+}$ 
($s=1$) ion, sites 1-7 are occupied by Cr$^{3+}$ ($s=3/2$) ions, 
and $ {\bf s}_9 \equiv {\bf s}_1 $. The first term 
accounts for the isotropic exchange interaction, while the second and third ones are the dominant axial contributions to the crystal-field and the intracluster 
dipole-dipole interactions, respectively. The last term represents the Zeeman coupling to an external 
magnetic field. 
The parameters entering the above Hamiltonian are determined by fitting the experiments performed with (ensembles of) single rings
(see Fig.\ref{HC}, for instance).
Intra-ring interactions are also responsible for the anomalies above few 
K in the supramolecular structures; the analisis of these features shows that the parameters are not affected by the intermolecular coupling introduced in the ring dimers and oligomers.
Then, low temperature anomalies are described at a microscopic level by considering the interaction of Cu spin centre with Ni and Cr spins of
each rings \cite {NanoNature}. Considering also the projection of the rings dipolar and crystal fields, the effective interaction can be written as:
\begin{equation}\label{effham}
\mathcal{H}= J^* {\bf S}^{Cr7Ni} \cdot {\bf S}^{Cu}+D_{ex}^* [2S^{Cr7Ni}_z{S}^{Cu}_z -S^{Cr7Ni}_x{S}^{Cu}_x
- S^{Cr7Ni}_y{S}^{Cu}_y]
\end{equation}
for each Cr$_7$Ni - Cu pair. The $J^*$  and $D_{ex}$ parameters are evaluated by simultaneously
fitting complementary experimental results \cite{NanoNature}.

\subsection{ understanding the magnetic coupling}

\begin{figure}[ptb]
\begin{center}
\includegraphics[width=12cm]{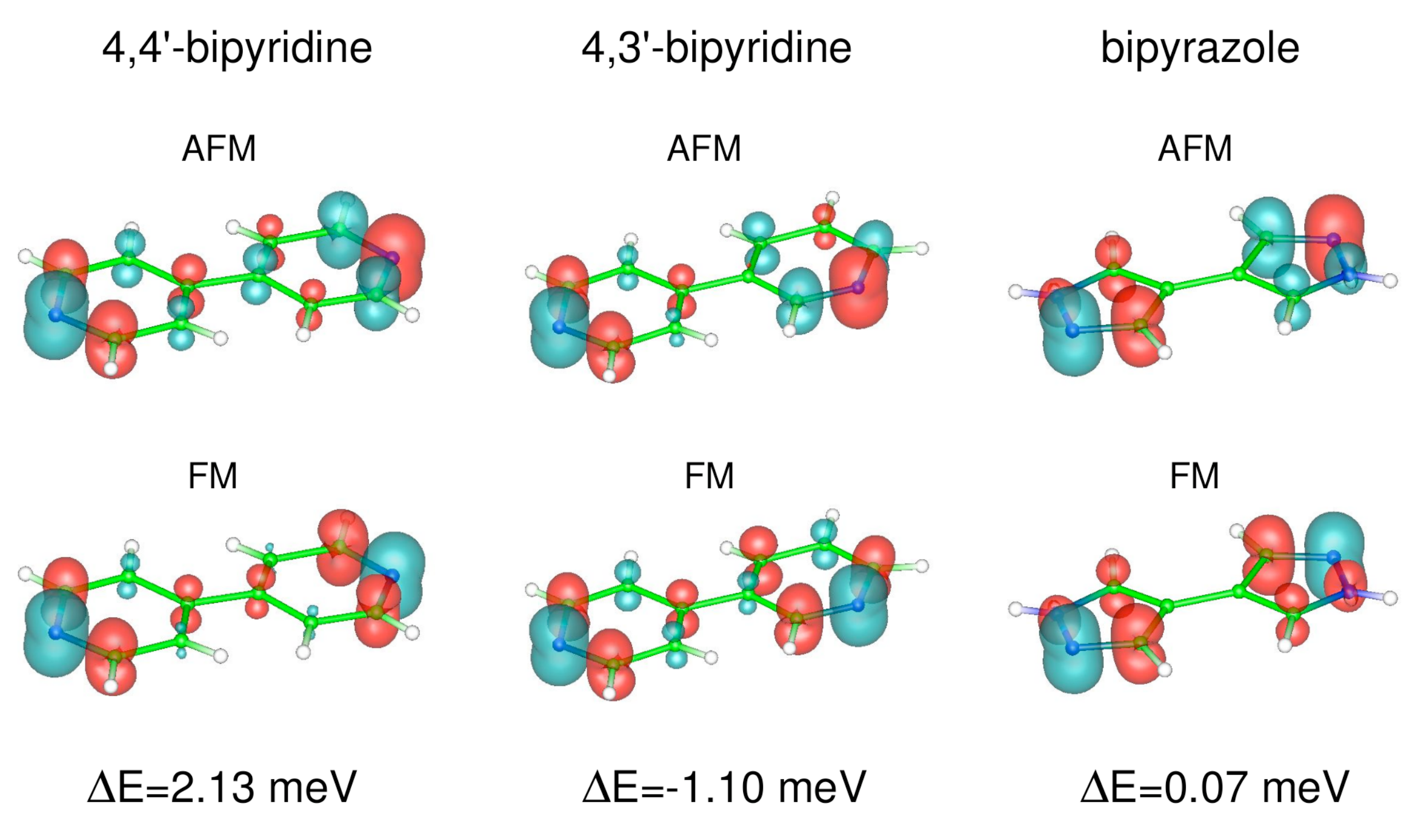}
\end{center}
\caption{Spin density isosurfaces for isovalue of +0.001 electrons/a.u. (blue
color), and -0.001 electrons/a.u. (red color), and FM-AFM energy splitting
for 4,4$^{\prime}$"-bipyridine, 4,3$^{\prime}$"-bipyridine and bipyrazole
bridges (see text for details). }
\label{spindensity}
\end{figure}

\vspace{1cm}
How the organic linkers actually transmit spin information is an interesting issue that may help in designing organic linkers and new experiments.
The series of [Cr$_7$Ni]$_2$ dimers discussed in a previous section is quite instructive from this point of view.
The linker in those cases belongs to heteroaromatic organic groups (C-based benzene-like rings containing one or more nitrogen)
that have been long studied and intensively used in the '80s and '90s in order to carry electronic and magnetic
interactions between active molecular units through long (nm) distances,
as compared to standard organic bridges as single O or F atoms, hydroxides or carboxylates that, conversely, 
work at atomic scale. Here the figure of merit, which discriminates between ``good'' and ``bad'' linker, is the level of
conjugation/delocalization of the electrons that carry the information. $p$ electrons are distributed over two types of orbitals,
the ones that bind the linker atoms together ($sp^3$ hybrids, with label $\sigma$ ) and $\pi$ electrons that occupies resonant
and delocalized bonds.  
Magnetic interaction is optimal when large overlap (both in space and in energy) between the spin polarized orbitals
of the magnetic centers and orbitals of the linker atom anchored to the magnetic centers is found; symmetry matching is also
important. In principle, both $\sigma$ and $\pi$ electrons can carry magnetic interactions,
although only with $\pi$ electrons delocalization is strong enough to drive this interactions
over long distances. Experimental observation of such interactions have been supported by numerical calculations,
mostly performed by H\"uckel (extended) molecular orbital methods.
Some general rules have been suggested in the literature:
an interesting observation is that spin polarization of $\pi$ electrons is found to proceed
with an oscillating character, moving from one atom to the other through aromatic groups. This results in
ferromagnetic or antiferromagnetic interaction between magnetic centers at the edges depending on where they
anchor \cite{cargillthompson1996,mccleverty1998}.  The strength of the interaction also obeys such alternation rule,
as discussed by Richardson and Taube \cite{richardson1983}, that has been also interpreted
as arising from a quantum interference over magnetic paths with different lenghts \cite{marvaud1993}.
As a matter of fact, alternation in spin and charge polarization through an aromatic linker
can be theoretically explained by superexchange mechanism discussed by McConnell \cite{mcconnell1963} or, 
alternatively, by resonant theories, as found by Longuet-Higgins \cite{longuet-higgins1950a}.
Another fact that has to be taken into account is that both occupied and unoccupied
orbitals can play a role, like in charge transport \cite{browne2006}.
Either $\pi$ occupied or unoccupied linker orbitals can be close in energy 
to the magnetic frontier orbitals, depending if the heteroaromatic linker is $\pi$ rich or $\pi$ poor,
thus either HOMO- or LUMO-driven magnetic superexchange interaction can be promoted. \\
In order to illustrate this mechanism we present \textit{ab-initio} DFT calculations on model bicycle organic linkers, namely
on bipyridine, both in the (4,4$^{\prime}$ and 4,3$^{\prime}$ configuration), and bipyrazole.
Calculations have been performed with the NWChem quantum chemistry package \cite{nwchem};
an Ahlrichs valence double zeta (VDZ) contracted gaussian basis set has been used in conjunction with the hybrid
B3LYP exchange and correlation functional.
Let us focus on the 4,4$^{\prime}$ bipyridine bridge depicted in Fig. \ref{spindensity}, 
and assume that the anchoring site is the N atom, more electronegative than C. 
Let us suppose also that the overlap between the magnetic frontier orbitals of the metal atoms
anchored to the N sites with the N orbitals is such that a small spin polarization of $\pm 0.1\mu_B$ is induced
on the N atoms (the signs refer to a ferro- or antiferromagnetic coupling between these two moments).
We impose such spin moment by means of the constrained DFT method as discussed in Ref. \cite{wu2005}.
We obtained spin polarized states in N with both $\sigma$ and $\pi$ characters, so that both symmetries can contribute to the
magnetic interaction, whereas most of the interaction is reasonably associated to the conjugated $\pi$ electron system.
In Fig.  \ref{spindensity} we plot the spin-polarized electron
density isosurfaces for isovalues of $\pm 0.001$ electrons/a.u. in case of
antiferromagnetic and ferromagnetic coupling between the N spin moments.
We can clearly see the alternation of spin polarization when moving from one atom to the next, 
following the bond paths. In order to demonstrate the rules discussed above, we plot in Fig.  \ref{spindensity} the spin densities
also for the 4,3$^{\prime}$ bipyridine and for bipyrazole organic bridges, always assuming that the metal
is anchored to N sites, and that a spin moment of $\pm0.1 \mu_B$ is transferred to N atoms.
Changing from 4,4$^{\prime}$ bipyridine to 4,3$^{\prime}$ bipyridine, 
optimal coupling is attained when the spin polarization on the two N atoms has the same sign, i.e. magnetic
centers are more favourable ferromagnetically coupled, as compared to the antiferromagnetic coupling attained 
for 4,4$^{\prime}$ bipyridine, as demonstrated by the larger spin polarization of the inner C atoms at the frontier
between the two pyridines. In bipyrazole, the spin densities, for both the antiferro- and
ferromagnetic states, do not reach the inner region and interference between the spin paths
hinder magnetic interaction between the two sides of the bridge. In Fig. \ref{spindensity} we report also the total energy
difference between the antiferromagnetic and ferromagnetic states, which indicates clearly how the size and the sign
of the coupling are completely consistent with the reasonings above.

In order to predict the behaviour of real dimeric complexes, the full systems, not only the bridge, have to be simulated.
We analyze three supramolecular dimers of [Cr$_7$Ni]$_2$, which are characterized by identical
magnetic molecular centers, two \textit{purple} Cr$_7$Ni, but three different organic linkers, i.e. bipyridine, bipyrazole and
bipyridylethylene \cite{tobepublished}.
Magnetic frontier orbitals are supplied by Ni ions, and anchoring sites in the linkers are always
N atoms. Ni (II) ions have nominally a 3$d^8$ electronic configurations, so that only $d$ orbitals with \textit{e$_g$} symmetry
are spin-polarized. Although this in principle should imply that only $\sigma$ orbitals of the extended molecule are responsible
of the spin interaction between the two rings, we observe that polarization of both $\sigma$ and $\pi$ orbitals in the linker
is present. $\sigma$ polarization retains values only for the C atoms in the vicinity of the N atoms, while for more distant C atoms,
only $\pi$ orbitals attain a (small) spin-polarization.
Intramolecular Heisenberg $J^*$ parameters, which quantify the interaction between the two Cr$_7$Ni molecules 
can be estimated from several experimental methods (as described in previous section) or calculated by total energy
(obtained by means of, e.g., DFT-B3LYP calculations) difference methods. 
Here the microscopic interaction arises through the organic bridges between the two Ni spin moments, so that the
relevant microscopic Hamiltonian is given by

\begin{equation}
H =  J^{*} S_{Ni^1} \cdot S_{Ni^2}  \,\, ,
\end{equation}

\noindent where the labels $1$ and $2$  indicate the Ni atoms belonging to different rings, and $S_{Ni^1}=
S_{Ni^2}=1$; $J^{*}$ is then given by 1/2 of the total energy difference between the singlet and triplet
states of the (Cr$_7$Ni)$_2$ dimer, and positive values are relative to a preferred antiferromagnetic coupling between the rings,
i.e. a singlet spin ground state. The calculated  $J^{*}$ values evidence a stronger magnetic coupling for bipyridine-bridged dimers ($J^{*}$=0.021meV) while for bipyrazole- ($J^{*}$=0.004meV) or bipyridylethylene- ($J^{*}$=0.002meV) bridged
dimers interaction is sensibly smaller, in agreement with specific heat measurements that provide an estimate of the energy gap between the singlet and the barycenter of the triplet of 0.009meV for bipyrazole-bridged and weaker ones (0.005meV and 0.004meV) for bipyrazole- and bipyridylethylene-bridged respectively.
In case of bipyrazole, as anticipated above, quantum interference between the two paths seems to be
the responsible for the small J, despite the fact that the two Ni centers are closer to each other, as compared to bipyridine,
because of the shorter length of bipyrazole. In case of the bipyridylethylene bridge (not shown), the larger number of bonds that such
interaction should travel through plays a role, so that only a small fraction of spin-polarization survives in the two facing C atoms in the 
center of the bridge \cite{tobepublished}.
These findings pave the way for a whole series of possible experimental investigations,  by systematically varying the organic bridges
and the magnetic frontier atoms, in order to tune and choose the appropriate magnetic coupling for entanglement. 
The reasonings above, that have been derived in dimeric complexes, apply as well for trimeric or tetrameric systems,
as the ones described in previous section; in these cases, some additional difficulties might be represented by the
many possible and simultaneous interaction paths, a circumstance that might prevent to prefigure the magnetic properties
of the systems by simple general conjectures, requiring that a full theoretical characterization has to be necessarily carried out.

\subsection{Switchable molecular links.}
Although switchability is not mandatory for entanglement and spin manipulation can also be obtained between
permanently coupled spins \cite{QC2}, we briefly discuss switchable organic linkers. We focus on three different
switching mechanisms, namely a mechanical, an electric field-induced  and a photocromic one.
Critical issues like the switching rate or preservation of coherence are far beyond current discussion but,
at the end, they will constitute possible bottlenecks for switchable linkers.\\
Transport through aromatic bicycles linkers have been demonstrated to depend on the structural conformation of the linker
\cite{venkataraman2006}. In the recent work of Quek et al. \cite{quek2009}, it has been demonstrated that
transport properties of  bipyridine-based molecular junction are modified by elonging or compressing the junction;
theoretical investigations have helped in
attributing this finding to modification in the internal angles of the linkers, and in bond lengths and angles
defining the pyridine-gold contact geometry. As discussed above, magnetic properties of supramolecular systems depend
similarly on the structural confomation of the linker and of the linker-molecule contacts, leading to the idea of mechanical switching
of the magnetic interactions. Another approach on the same line is the use of molecular shuttles \cite{molmotor} as possible switches.\\
Another possibility is to exploit a local electric field to rearrange the molecular orbitals and to disrupt/enhance energy
matching between orbitals of the linker and of the magnetic center.
As discussed by Diefenbach and Kim \cite{diefenbach2007} one can exploit the different spatial distribution
of the molecular orbitals in the linker, and more precisely their different polarizability; upon the application of a
(strong enough) electric field, the energetic order of the different orbitals might be modified, since second-order Stark
response might be very different for the different orbitals. In case of low lying excited spin states, crossings between
excited and ground states can be induced, which means that a different magnetic ground state can be fostered, that is,
magnetic swithing to on/off states can be achieved. 
Switchability of the linker is often used in other solid state systems, like, for instance, quantum dots. In this respect,
an interesting case was proposed considering a molecular poly-oxometallate [PMo$_{12}$O$_{40}$(VO)$_2$]$^{q-}$
consisting of two (VO)$^{+2}$ moieties with spin 1/2 separated by Mo$_{12}$ cage \cite{POM}.
The cage may have different valence states and it can therefore be charged providing a switcheabe link between
the two S=1/2 spins. The implementation of a square-root-of-swap gate has been proposed  \cite{POM} and
experimental work is in progress in this direction.\\
The latter switching method is the one exploiting photoexcitation processes. 
Photocromic linkers belonging to the family of  diarylethenes \cite{matsuda2000}, undergoes reversible conformational changes
upon irradiation in the visible or ultraviolet frequency range. They are optimal candidates because of their
resistance, rapid photo-response (in the range of picoseconds), and thermal stability of the two different isomers
(up to 100 $^{\circ}$C). Bonds form or brake, and conjugation is suppressed or enhanced, upon irradiation when moving
from one isomer to the other; magnetic interaction paths efficiency can be in this way controlled by photoirradiation.
These molecular switches are excellent candidates for large-scale integration too, since photocromic complexes have been
demostrated to react both in solution and in the crystalline phase, and last but not least, to be compatible 
with coordination-driven self-assembly synthetic approaches.

\subsection{quantify and measuring entanglement in molecular spin clusters}

The existence of a magnetic coupling between the molecular spin clusters doesn't guarantee per se that these are in an entangled state,  but its form plays a crucial role in the controlled generation of entanglement. 
Therefore, the high degree of flexibility with which such coupling can be engineered through supramolecular chemistry represents a fundamental resource. To illustrate how the main concepts apply to the molecular systems, we consider two different approaches to the generation of entanglement in coupled Cr$_7$Ni rings, the first one based on equilibrium states $ \rho $ at low temperature, the second one on coherent manipulation of the system state by electron-paramagnetic resonance (EPR) pulses. 
Being our interest focused on entanglement between the total spins of the nanomagnets that compose the supramolecular structure, we shall refer to the spin Hamiltonian approach.
Here, if the intermolecular interaction is small as compared to the 
intramolecular exchange coupling $J$, it can be treated at a perturbative level, and mapped
onto an effective Hamiltonian $ \mathcal{H}_{eff}^{AB} $ 
that depends only on the total spins $ {\bf S}_\alpha $ ($ \alpha = A , B , \dots $) of the coupled molecules.
Both the expression of $ \mathcal{H}_{eff}^{AB} $ and the values of the effective parameters
are deduced from the underlying microscopic model.

In order for the equilbrium density matrix to be entangled, one typically needs 
an intermolecular coupling Hamiltonian $ \mathcal{H}_{eff}^{AB} $ with a non factorizable 
ground state, and such that the energy separation from the first excited state is significantly larger than
the lowest temperature at which relevant experiments can be performed. 
In the case of the (Cr$_7$Ni)$_2$ dimer ($ S_A = S_B = 1/2 $), the former condition 
can be achieved if the dominant term in the coupling Hamiltonian is an antiferromagnetic exchange interaction.
Anisotropic intramolecular interactions give rise to additional effective terms, resulting in the following Hamiltonian:
$ \mathcal{H}_{eff}^{AB} = (J_{AB}-D_{AB}) {\bf S}_A \cdot {\bf S}_B + 3 D_{AB} S^A_z S^B_z $. 
For temperatures comparable with $J_{AB}$, the equilibrium state $ \rho $ includes 
contributions from all four lowest eigenstates $ | S , M \rangle $ (being $S$ 
and $M$ the total spin and its projection along $z$, orthogonal to the plane of 
the molecules), with Boltzmann probabilities $P^S_M$. 
The entanglement between two 1/2 spins can be quantified by the concurrence 
($\mathcal{C}$), whose value ranges from 0 for a factorizable $ \rho $ to 1 
for maximally entangled states \cite{fazio}. 
In the present case, 
the expression of 
$ \mathcal{C} $ corresponding to the equilibrium state reads:
\begin{equation}
\mathcal{C} ( P^S_M ) = \left\{ \begin{array}{ll} 
\max \{ | P^1_0 - P^0_0 | - 2\sqrt{P^1_1 P^1_{-1}}, 0\} 
  & {\rm  for} \max\{ P^1_0 , P^0_0 \} > \sqrt{P^1_{-1} P^1_1} \\
0 & {\rm  otherwise}  
\end{array}\right.
\end{equation}
In the presence of a magnetic field applied along the ring axis, the 
expression of the concurrence reads:
\begin{equation}
\mathcal{C}(\rho_{eq}^{AB}) =
\frac{ 1 - e^{-\frac{J_{AB}}{k_BT}} \left( e^{\frac{D_{AB}}{k_BT}} + 2e^{-\frac{D_{AB}}{2k_BT}} \right) }
     { 1 + e^{-\frac{J_{AB}}{k_BT}} \left[ e^{\frac{D_{AB}}{k_BT}} + 2e^{-\frac{D_{AB}}{2k_BT}} \cosh \left(\frac{\bar{g}_{zz}\mu_B B}{k_BT}\right) \right] } ,
\end{equation}
where $ \bar{g}_{zz} $ is the $z$ component of the effective g factor in the 
ground state $ S = 1/2 $ doublet of the Cr$_7$Ni ring.
According to this expression, that holds as long as $ | S, M \rangle $ are the 
dimer eigenstates, the molecular spin clusters $A$ and $B$ are entangled if the occupation 
of either $ |0,0\rangle $ or $ |1,0\rangle $ is sufficiently larger than all 
the others.
In particular, in the limit $ k_B T \ll ( J_{AB} - D_{AB} ) $, the equilibrium state 
tends to the singlet ground state and $ \mathcal{C} \simeq 1 $.
Therefore, the larger $J_{AB}$, the wider the temperature range in which thermal 
entanglement persists. In the present case,
the range of desirable values of $J_{AB}$ is however bounded from above by the 
characteristic energy of the intramolecular spin excitations. 
If this condition is not fulfilled, each nanomagnet within the dimer can no longer be regarded as effective two-level systems, for 
intramolecular excitations corresponding to higher spin multiplets enter the composition of the dimer lowest eigenstates. 
The concurrence exponentially decreases with the magnetic field, that we assume
for simplicity oriented along $z$. In fact, the field energetically favours the 
factorizable ferromagnetic state ($M=1$) and reduces the occupation of the 
singlet state. At zero temperature, an abrupt transition takes place as a 
function of $B$, at the level crossing between $ | 1, 1 \rangle $ and 
$ | 0, 0 \rangle $.
In general, the concurrence cannot be easily expressed in terms of observable
quantities. Its evaluation requires the knowledge of the system density matrix,
that is either derived experimentally from quantum state tomography or 
indirectly through the determination and diagonalization of the system 
Hamiltonian. The latter approach is in general viable in the case of a few
coupled molecular spin clusters, where a detailed knowledge of the system Hamiltonian can 
be achieved by simulating a number of experimental techniques, including 
specific heat, torque magnetometry, inelastic neutron scattering, electron 
paramagnetic resonance as discussed in a previous paragraph.\\

The demonstration of quantum entanglement, however, can also be directly 
derived from experiments, without requiring the knowledge of the system 
state. This can be done by using specific operators - the so-called {\it entanglement
witnesses} - whose expectation value is always positive if the state $ \rho $
is factorizable. It is quite remarkable that some of these entanglement 
witnesses coincide with well known magnetic observables, such as energy or magnetic susceptibility $ \chi = dM / dB $.
In particular, the magnetic susceptibility of $N$ spins $s$, averaged over three orthogonal spatial directions,
is always larger than a threshold value if their equilibrium state 
$ \rho $ is factorizable: $ \sum_\kappa \chi_\kappa > N s / k_B T $ \cite{Wie}.
This should not be surprising, since magnetic susceptibility is proportional to the variance of the magnetization,
and thus it may actually quantify spin-spin correlation.
The advantage in the use of this criterion consists in the fact that it 
doesn't require the knowledge of the system Hamiltonian, provided that this 
commutes with the Zeeman terms corresponding to the three orthogonal 
orientations of the magnetic field $ \beta = x,y,z $.
As already mentioned, in the case of the (Cr$_7$Ni)$_2$ dimer, the effective 
Hamiltonian includes, besides the dominant Heisenberg interaction, smaller 
anisotropic terms, due to which the above commutation relations are not 
fulfilled. This might in principle result in small differences between the 
magnetic susceptibility and the entanglement witness
$ \bar{\chi}_{EW} \equiv 
\sum_\beta [ \sum_{\alpha , \beta} \langle S_z^\alpha S_z^\beta \rangle - 
\langle \sum_\alpha S_z^\alpha \rangle ]$. Such difference is however negligibly 
small if $ D_{AB} $ is small compared to $ J_{AB} $ and to the temperature (see Fig. \ref{fig7}).
Magnetic susceptibility $\chi$ was used as entanglement witness in the case of Cr$_7$Ni dimers \cite{ent09}.
In figure \ref{entwit2} the product $\chi$T is plotted vs temperature and compared with the expected threshold.
In fact, in this system the ratio $ J_{AB} / D_{AB} \simeq 4 $ is large enough
to make the difference between the magnetic susceptibility and the entanglement 
witness negligible. \\

\begin{figure}[ptb]
\begin{center}
\includegraphics[width=15cm]{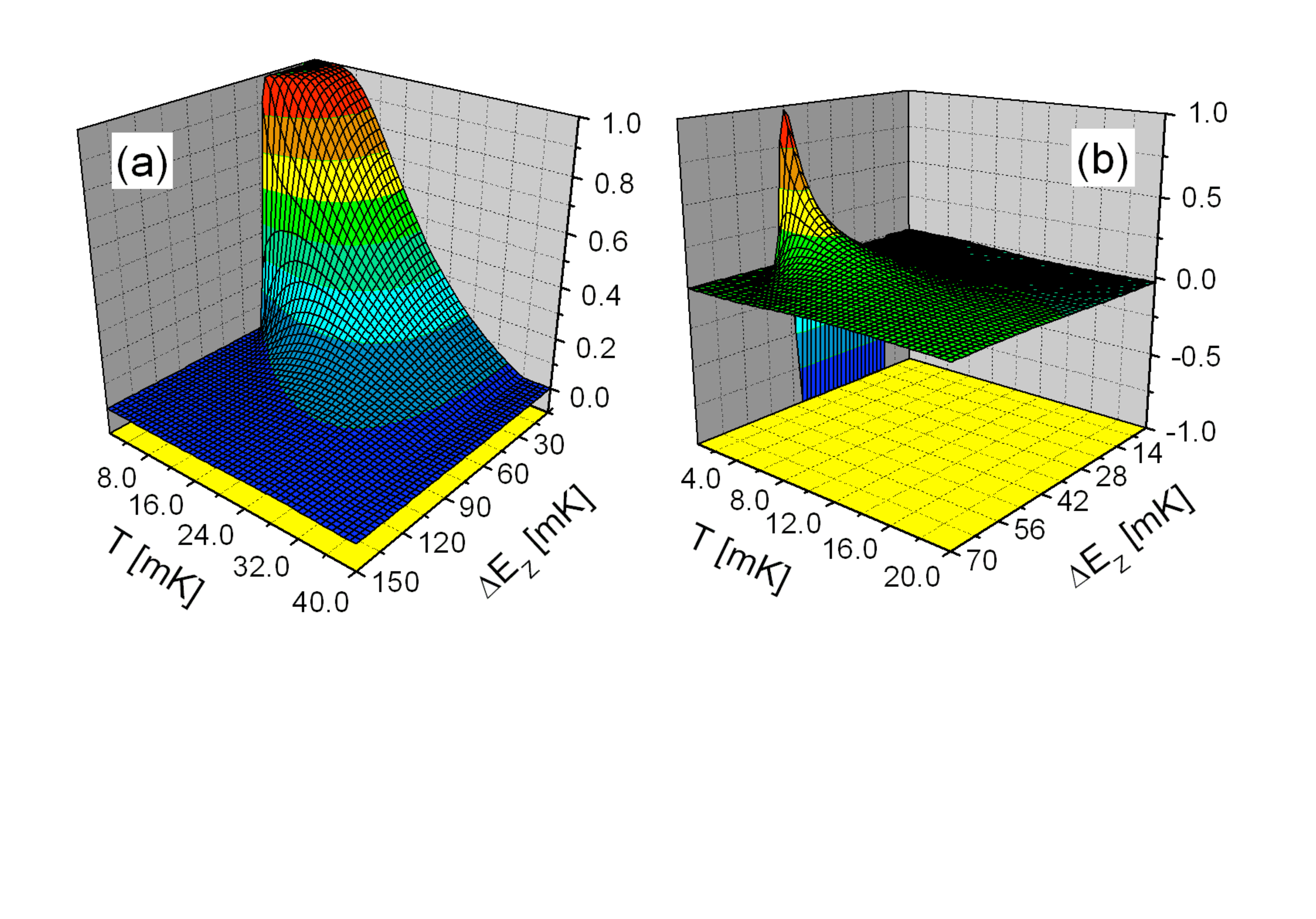}
\end{center}
\caption{(a) Concurrence of the ring dimer as a function of
temperature 
and of the Zeeman splitting induced by the applied magnetic field. The
values of the physical parameters entering the effective Hamiltonian 
$ \mathcal{H}^{AB}_{eff}$ are: $ J_{AB} = 40\, $mK and $ D_{AB} =
10\,$mK.
(b) Difference between concurrence in the presence of anisotropy 
and concurrence without anisotropy ($ D_{AB} = 0$). }
\label{entwit}
\end{figure}

\begin{figure}[ptb]
\begin{center}
\includegraphics[width=15cm]{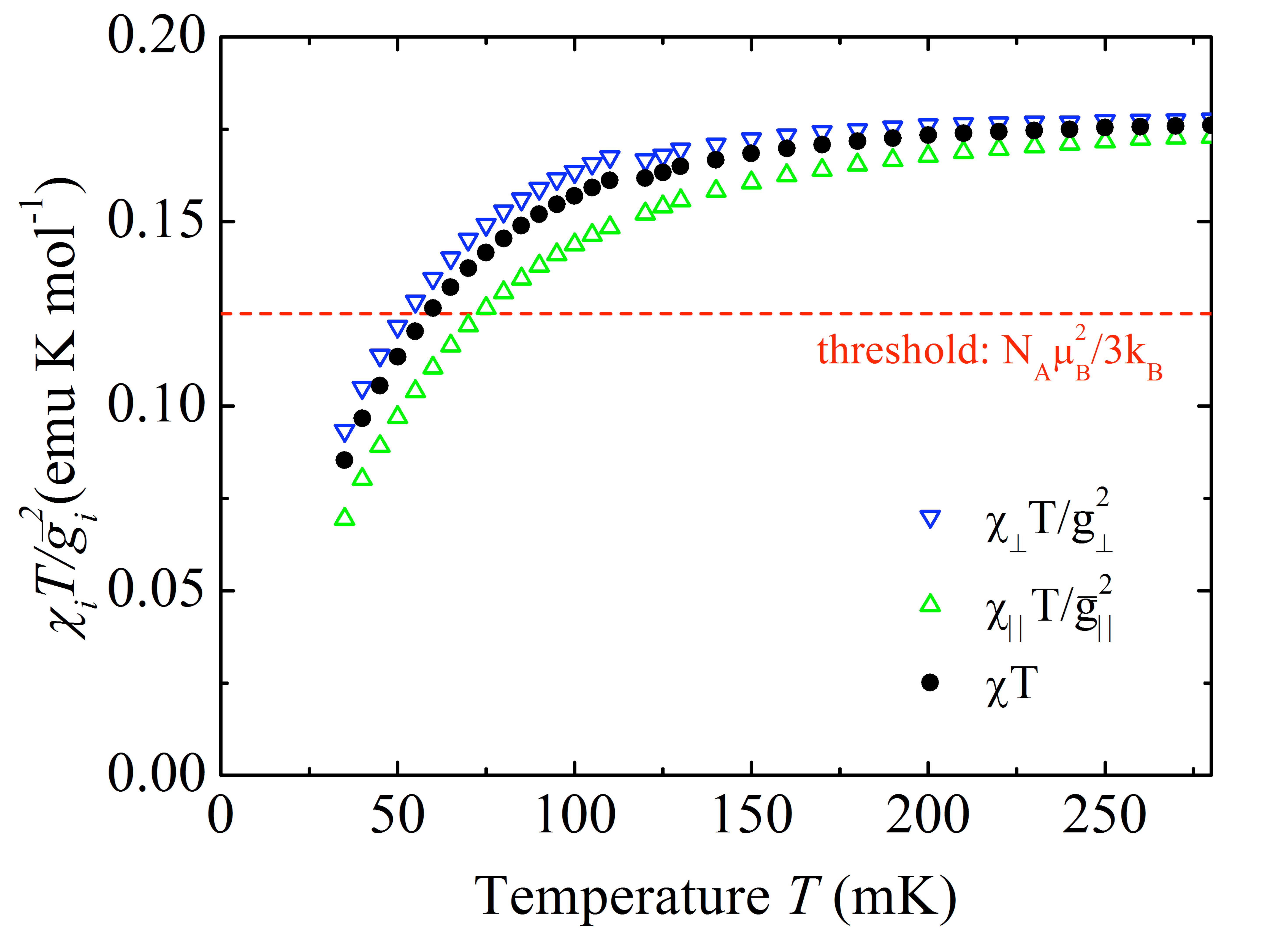}
\end{center}
\caption{Magnetic susceptibility $\chi$ used as entanglement witness in the case of Cr$_7$Ni dimers. Temperature dependence of the measured $\chi$T product (triangles). $ \chi_\perp $ (blue) is 
the component perpendicular to the largest surface of the crystal; this direction
forms on average an angle of $16^\circ$ with the $z$-axis, perpendicular to the 
rings plane. $ \chi_\parallel$ (green) refers to the directions parallel to the crystal
plane; rotation of magnetic field within this plane does not evidence changes in 
the magnetic response.
The average $ ( \chi_\perp + 2 \chi_\parallel ) / 3 $ (black dots) is compared with 
the threshold for a mole of dimers, $ N_A \mu_B^2 / 3 k_B $, in order to identify 
the temperature range (T$\leq$50mK) where the two rings are entangled \cite{ent09}.}
\label{entwit2}
\end{figure}

An alternative approach to the generation of entangled states is represented by 
the application of suitable EPR pulse sequences to an initially unentangled state. 
Broadly speaking, this requires the implementation of a conditional dynamics, 
where the effect produced by a given pulse sequence of a (target) nanomagnet 
$A$ depends non-trivially on the state of a (control) nanomagnet $B$. 
In the case where the dimer consists of two identical and equally oriented 
molecular spin clusters, limitations arise from the impossibility of individually addressing 
$A$ and $B$. In fact, it's easy to verify that an effective Hamiltonian such as 
$ \mathcal{H}_{eff} = \mathcal{H}_{eff}^{AB} + \sum_{\alpha = A,B} {\bf B} \cdot 
\bf{g}_\alpha \cdot {\bf S}_\alpha $, with $ {\bf g}_A = {\bf g}_B = {\rm diag }
(g_\perp , g_\perp , g_\parallel ) $ doesn't allow to generate an entangled state 
such as $ | S, 0 \rangle $, starting from a factorized one such as $ | 1 , \pm 1 
\rangle $. These limitations can be overcome in the case of an asymmetric system,
where the coupling of the two effective spins $A$ and $B$ with the magnetic field
are different, due either to the different chemical composition of the two 
molecular spin clusters or to their different spatial orientation, combined with the
anisotropy of the $g$ tensor ($ g_\parallel \neq g_\perp $). Alternatively, the 
asymmetry of the intermolecular coupling can be exploited, such as that between 
the green and the purple derivatives of Cr$_7$Ni \cite{Timco}. 

Analogous features allow the controlled generation of entangled states in 
tripartite systems. The (Cr$_7$Ni)-Cu-(Cr$_7$Ni) molecule, for example,
behaves as a system of three effective 1/2 spins ($ S_A = S_B = S_{Cu} = 1/2 $) \cite{NanoNature}.
Entanglement between three parties can manifest itself in fundamentally different
forms. In fact, two classes of equivalence have been defined, whose prototypical 
states are the so-called GHZ and W states, respectively. 
The GHZ states, whose expression in 
the $ | M_A, M_B, M_C \rangle $ bases reads:
$ | \Psi_{GHZ} \rangle = ( | 1/2, 1/2, 1/2 \rangle + |-1/2,-1/2,-1/2 \rangle ) / \sqrt{2} $, 
maximize the genuinely tripartite entanglement, i.e. the one that cannot be 
reduced to pairwise correlations. 
The expression of the W states reads instead:
$ | \Psi_{W} \rangle = ( | 1/2, 1/2,-1/2 \rangle + | 1/2,-1/2, 1/2 \rangle 
                       + |-1/2, 1/2, 1/2 \rangle ) / \sqrt{3} $, 
and coincides with that of the $ | S, M \rangle = | 3/2 , 1/2 \rangle $.
In order for the controlled generation of both $ | \Psi_{GHZ} \rangle $ and 
$ | \Psi_{W} \rangle $ to be possible, by applying suitable pulse sequences to 
an initial ferromagnetic state $ | 3/2 , 3/2 \rangle $, the degeneracy between 
the two transitions $ | \Delta M | = 1 $ within the $ S = 3/2 $ quadruplet needs 
to be broken. This is indeed the case for the (Cr$_7$Ni)-Cu-(Cr$_7$Ni) 
system, thanks to the anisotropic terms in the effective Hamiltonian 
(see Eq. \ref{effham})
and to the resulting zero-field splittings.

It's finally interesting to note that quantum correlations can also be present 
in the equilibrium state of the tripartite system described by the above 
effective Hamiltonian $ \mathcal{H}_{eff}^{AC} + \mathcal{H}_{eff}^{BC} $, for 
suitable values of the parameters $ J_{AC}=J_{BC} $ and $ D_{AC} = D_{BC} $. 
If the inter-ring interaction is dominated by the exchange term ($ J_{AC} \gg 
D_{AC} $), the anisotropy can be perturbatively included in first order, and 
the system eigenstates coincide with the vectors $ | S_{AB}, S, M \rangle $
($ {\bf S}_{AB} = {\bf S}_{A} + {\bf S}_{B} $). 
In the case of a ferromagnetic coupling ($ J_{AC} < 0 $), the density matrix 
in the low-temperature limit ($ k_B T \ll | J_{AC} | $) is given by a statistical
mixture of the $S=3/2$ eigenstates. If $ D_{AC} < 0 $ \cite{NanoNature}, 
the three pairs of subsystems are all unentangled. 
In the case of an antiferromagnetic coupling between the rings, the ground state
coincides with the state 
$ | S_{AB} = 1, S=1/2, M \rangle = (|1/2,-1/2,1/2 \rangle + 
|-1/2,1/2,1/2 \rangle -2|1/2,1/2,-1/2 \rangle ) / \sqrt{6} $.
If the tripartite system is cooled down to this state ($ k_BT \ll J_{AC}, 
g\mu_B B$), each subsystem is entangled with the other two.
In fact, the reduced density matrix of any two subsystems is real and takes the 
form:
\begin{displaymath}
\rho_{red}^{\alpha\beta} = \left(
\begin{array}{cccc}
\rho_{11} & 0 & 0 & 0 \\
0 & \rho_{22} & \rho_{23} & 0 \\
0 & \rho_{23} & \rho_{33} & 0 \\
0 & 0 & 0 & \rho_{44} 
\end{array}
\right) ,
\end{displaymath}
where we refer to the basis 
$ \{ | 1/2, 1/2\rangle , | 1/2,-1/2\rangle , 
     |-1/2, 1/2\rangle , |-1/2,-1/2\rangle  \} $ 
and 
$ \alpha, \beta = A, B, C $.
For $ \alpha\beta = AB $, the above matrix elements are: 
$\rho_{11}=2/3$, $\rho_{22}=\rho_{33}=\rho_{23}=1/6$, and $\rho_{44}=0$.
The resulting entanglement between the two rings is given by 
$ \mathcal{C} (\rho_{red}^{AB}) = 1/6 $.
Each ring is also entangled with the Cu ion. In fact, for $\alpha\beta = AC$,
the matrix elements are: 
$\rho_{11}=1/6$, $\rho_{23}=-1/3$, $\rho_{22}=2/3$, $\rho_{22}=1/6$, and 
$\rho_{44}=0$.
This results in a finite concurrence, namely 
$ \mathcal{C} (\rho_{red}^{AC}) = 2/3 $.

\begin{figure}[ptb]
\begin{center}
\includegraphics[width=15cm]{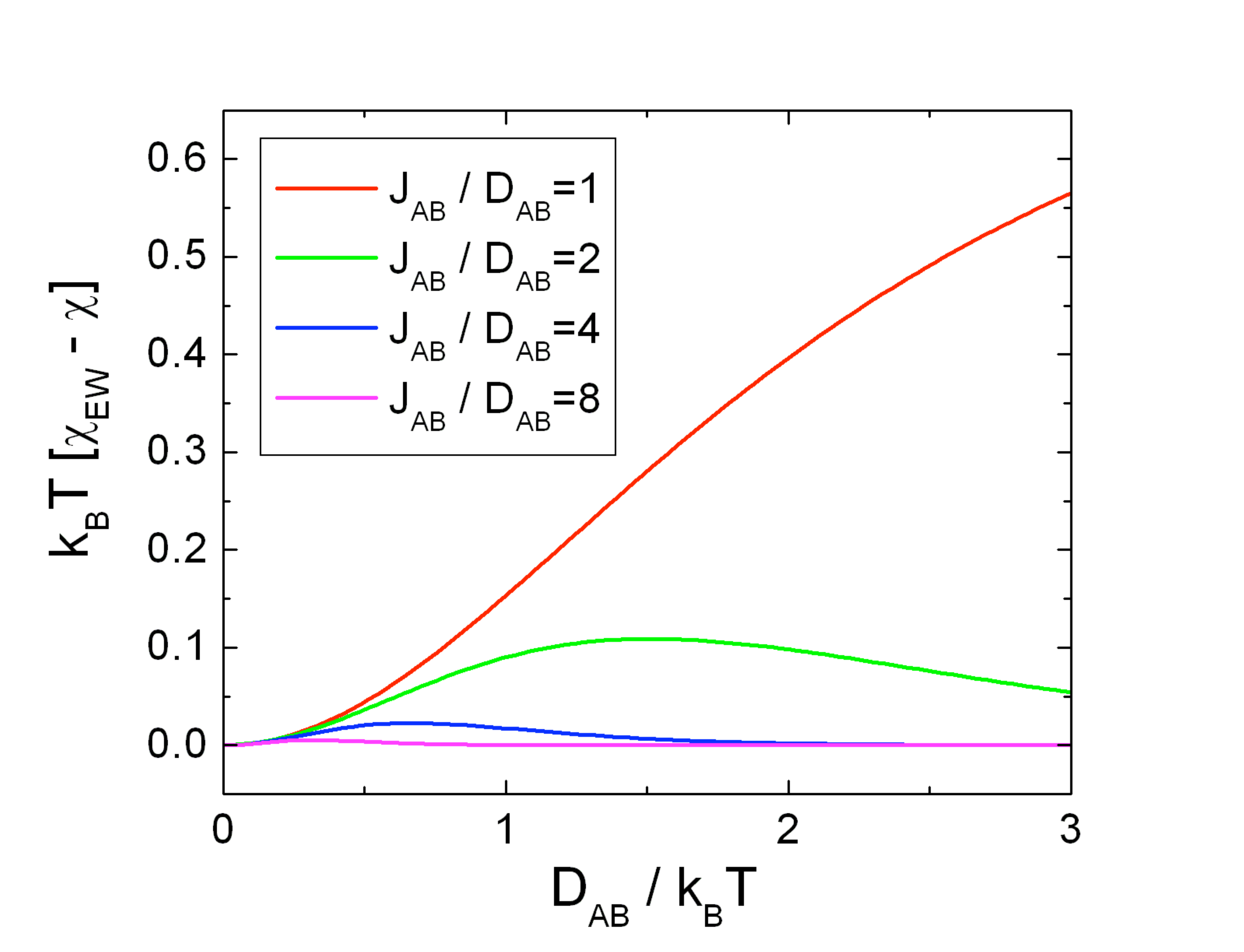}
\end{center}
\caption{Difference between the entanglement witness $ \chi_{EW} $ and
the magnetic susceptibility for the ring dimer, derived from the
effective Hamiltonian $ \mathcal{H}_{eff}^{AB} $ in the limit $ B \rightarrow 0 $.}
\label{fig7}
\end{figure}

\subsection{conclusions and perspectives.}

A quick look at the list of works cited here below tell us that entanglement in supramolecular systems has just
appeared as possible emerging topic, but the earliest results show great potentialities.
It is clear that advancements in this field may arrive only from the combined effort of chemists,
experimentalists and theoreticians.\\
From synthetic point of view, the list of suitable molecular building blocks and of organic ligands
working as efficient linkers is - if not infinite - certainly very long. We discussed the reasons why
molecular Cr$_7$Ni rings on one side and heteroaromatic ligands on the other side represent a
very good starting point to build weakly interacting molecular complexes. The combination of the
two (i.e. molecule + linker) is just limited by the rules of coordination chemistry, that may well
bring to several interesting cases.\\
Experiments to characterize systems are certainly not routine but quite accessible. The range of molecular
energies indeed spans between 0.01 to 10K that correspond to the energy of one electron in magnetic
field up to 10 Teslas and frequencies ranging from 0.01 to 20 cm$^{-1}$, that is microwaves with
low wavenumbers.  Molecular spin clusters also represent an ideal test bed to perform experiments
targeted at directly probing and quantify entanglement in spin systems. Here we have just mentioned
the use of magnetic susceptibility, independently measured along its three components, as entanglement
witness but other quantities, like specific heat or neutron scattering, may well do this job.
In the next future it will be certainly interesting to use pulsed electron spin resonance to address
selectively molecular subensembles. Here the possibility of spectroscopically discern different molecules
will be certainly of interest. Design of specific pulse sequences will lead to implement quantum algorithms.\\
From the theoretical point of view, finite arrays of molecular spins are very appealing to develop models.
Here one may wonder which conditions (forms of spin hamiltonian, values of the spin S$\neq$1/2,
number of spin centers,  etc.) maximize/minimize entanglement. As mesoscopic systems, molecular spin
clusters are paradigmatic cases to study crossover between quantum and classical behavior.
In particular it will be very instructive to study the role of decoherence mechanisms in details.

\subsection{acknowledgements}
We are indebted to Dr. Grigore Timco and Prof. Richerd Winpenny (University of Manchester, UK) for sharing and discussing their results with us. Magnetization measurements were taken by microSQUID in collaboration with Dr. Wolfgang Wernsdorfer in Grenoble (F). We thank Dr. Alberto Ghirri and Christian Cervetti (CNR and University of Modena, I) for contributing at low temperature characterization and Dr. S. Carretta, Prof. P. Santini and Prof. G. Amoretti (Univerity of Parma, I) for stimulating discussion.
This work is partially supported by the European projcet FP7-ICT FET Open "MolSpinQIP" project, contract N.211284.

\section*{References}



\end{document}